\documentclass[fleqn,usenatbib]{mnras}

\usepackage{newtxtext,newtxmath}
\usepackage[dvipsnames]{xcolor}

\usepackage[T1]{fontenc}
\usepackage{ae,aecompl}
\usepackage{color}

\usepackage{graphicx}	
\usepackage{amsmath}	
\usepackage{amssymb}	

\title[NGC 1313 X-1 time lags]{Discovery of a soft X-ray lag in the Ultraluminous X-ray Source NGC~1313 X-1}

\author[E. Kara et al.]{E. Kara$^{1}$\thanks{E-mail: ekara@mit.edu}, C.~Pinto$^{2,3}$, D.J.~Walton$^{4}$, W.N. Alston$^{4}$, M. Bachetti$^{5}$, D. Barret$^{6}$, \newauthor  M. Brightman$^{7}$, C.R. Canizares$^{1}$, H.P. Earnshaw$^{7}$,  A.C.~Fabian$^{4}$, F. F\"urst$^{8}$,   P. Kosec$^{4}$, \newauthor M.J.~Middleton$^{9}$,  T.P.~Roberts$^{10}$, 
 R. Soria$^{11,12,13}$, L. Tao$^{7}$,  N.A.~Webb$^{14}$
\\
$^{1}$MIT Kavli Institute for Astrophysics and Space Research, Cambridge, MA 02139, USA\\
$^{2}$ESTEC/ESA, Keplerlaan 1, 2201AZ Noordwijk, The Netherlands \\
$^{3}$INAF - IASF Palermo, Via U. La Malfa 153, I-90146 Palermo, Italy\\
$^{4}$Institute of Astronomy, Madingley Road, CB3 0HA Cambridge, United Kingdom \\
$^{5}$INAF-Osservatorio Astronomico di Cagliari, via della Scienza 5, I-09047 Selargius, Italy\\
$^{6}$Universit\'e de Toulouse, CNRS, Institut de Recherche en Astrophysique et Plan\'etologie, 9 Avenue du colonel Roche, BP 44346, 31028 Toulouse Cedex 4, France\\
$^{7}$Cahill Center for Astronomy and Astrophysics, California Institute of Technology, 1216 East California Boulevard, Pasadena, CA 91125, USA\\
$^{8}$European Space Astronomy Centre (ESAC), Science Operations Departement, 28692 Villanueva de la Ca\~nada, Madrid, Spain\\
$^{9}$Department of Physics and Astronomy, University of Southampton, Highfield, Southampton SO17 1BJ, UK \\
$^{10}$Centre for Extragalactic Astronomy, Durham University, Dept of Physics, South Road, Durham DH1 3LE, UK \\
$^{11}$National Astronomical Observatories, Chinese Academy of Sciences, Beijing 100012, China\\
$^{12}$International Centre for Radio Astronomy Research, Curtin University, GPO Box U1987, Perth, WA 6845, Australia\\
$^{13}$Sydney Institute for Astronomy, School of Physics A28, The University of Sydney, Sydney, NSW 2006, Australia\\
$^{14}$CNRS, IRAP, 9 avenue du Colonel Roche, BP 44346, F-31028 Toulouse Cedex 4, France\\
}

\begin{document}
\label{firstpage}
\pagerange{\pageref{firstpage}--\pageref{lastpage}}
\maketitle

\begin{abstract}
Ultraluminous X-ray Sources (ULXs) provide a unique opportunities to probe the geometry and energetics of super-Eddington accretion. The radiative processes involved in super-Eddington accretion are not well understood, and so studying correlated variability between different energy bands can provide insights into the causal connection between different emitting regions. We present a spectral-timing analysis of NGC~1313~X-1 from a recent XMM-Newton campaign. The spectra can be decomposed into two thermal-like components, the hotter of which may originate from the inner accretion disc, and the cooler from an optically thick outflow. We find correlated variability between hard (2--10~keV) and soft (0.3--2~keV) bands on kilosecond timescales, and find a soft lag of $\sim150$ seconds. The covariance spectrum suggests that emission contributing to the lags is largely associated with the hotter of the two thermal-like components, likely originating from the inner accretion flow. 
This is only the third ULX to exhibit soft lags.  The lags range over three orders of magnitude in amplitude, but all three are $\sim5$ to $\sim20$ percent of the corresponding characteristic variability timescales. If these soft lags can be understood in the context of a unified picture of ULXs, then lag timescales may provide constraints on the density and extent of radiatively-driven outflows.

\end{abstract}


\section{Introduction}

Ultraluminous X-ray Sources (ULXs) provide a unique opportunity to study super-Eddington accretion in the nearby universe (see \citealt{kaaret17} for a recent review). The nature of these luminous ($L_{X}>10^{39}$~erg/s), off-nuclear point sources has long been debated, with some claims that they were intermediate-mass black holes accreting at sub-Eddington rates (e.g. \citealt{Miller2004}), and others that they represent a population of stellar mass compact objects accreting at high rates (e.g. \citealt{King2001,Poutanen2007}). With the discovery of pulsations in several ULXs \citep{Bachetti2014, Furst2016,Israel2017a,Carpano2018,Sathyaprakash2019a,RodriguezCastillo2019}, it is now clear that it is possible, at least for neutron stars, to produce emission that appears up to 500 times the Eddington limit. 

Accretion flows in the super-Eddington regime are dominated by radiation pressure, which can cause massive outflows that reach mildly relativistic velocities (see e.g. simulations of \citealt{Takeuchi2013,jiang14,McKinney2015}). A super-Eddington accretion flow will therefore be geometrically and optically thick. While this basic geometry is agreed, major questions persist about the overall radiative efficiency and how much energy and mass can be carried out in the wind. Recently, \citet{Pinto2016nature} placed important constraints on these radiatively-driven outflows through the discovery of highly blueshifted absorption features in two ULXs, including NGC~1313~X-1, the focus of this letter. These absorption features have now been seen in several ULXs \citep{Pinto2017,Kosec2018a,Walton2016a}, including one that is a confirmed pulsar \citep{Kosec2018b}.

The broadband spectra of many ULXs also provided an early indication of their super-Eddington origin \citep{Gladstone2009}. They exhibit extremely soft X-ray spectra, with little hard X-ray emission above 15~keV \citep{Bachetti2013,walton13}. Hard X-ray emission up to $\sim 100$~keV is nearly ubiquitous in sub-Eddington active galactic nuclei (AGN) and accreting black hole X-ray binaries in our Galaxy, suggesting that ULXs are not in the same accretion regime. While clearly different from sub-Eddington accretors, a detailed understanding of the physical processes producing the observed radiation is still not yet well understood. ULX spectra in the 0.3--10~keV range can be modelled  phenomonologically as two thermal-like continua with temperatures of $\sim 0.2$~keV and a few keV \citep{stobbart06}. \citet{Poutanen2007} suggested that the hotter thermal component is produced by the inner thick disc (with some distortions due to inverse Compton upscattering by hot electrons in the inner accretion flow, or Compton downscattering by cooler electrons in the wind; \citealt{middleton15}). The lower temperature component originates further out around the spherization radius of the disc (i.e. where the flow becomes locally super-Eddington and a radiatively driven wind is launched; \citealt{SS1973,Poutanen2007}). The intrinsic inner disc emission can also thermalize the outflow, adding to the lower-temperature blackbody component.

This model can be extended to explain some of the short and long timescale variability properties seen in ULXs (e.g. \citealt{Middleton2011}). Regardless of the energy band, ULXs generally show less short-timescale variability relative to sub-Eddington AGN and black hole X-ray binaries \citep{heil09,Bachetti2013}. This could be the result of a supression of intrinsic, short-timescale variability (e.g. $\lesssim 1$Hz) as hotter inner disc emission gets scattered in a thick shroud of electrons \citep{middleton15,mushtukov19}. On longer timescales, inhomogeneities in the outflow (as seen in some hydrodynamical simulations of super-Eddington flows; \citealt{Takeuchi2013}), could cause additional variability, as photons scatter off or are absorbed and re-emitted by the clumpy outflow \citep{Middleton2011}.

Understanding the causal connection between spectral components is essential for understanding the ULX emission mechanism, and to this aim, several authors have searched for correlated variability and time lags in these sources. Unfortunately, due to the low count rate and low variability of most ULXs this is no easy feat, but there are currently two ULXs that show time lags. \citet{heil10} first found evidence for a soft-band lag in the ULX NGC~5408 X-1 at frequencies of $\sim 10$~mHz (confirmed by \citealt{demarco13,hernandez-garcia15}). Recently, another soft lag was detected in the very soft ULX NGC~55 X-1 at much lower frequencies, $f\sim0.1$~mHz \citep{Pinto2017}. We emphasize that this lag was two orders of magnitude larger than the one found in NGC~5408~X-1, and so while both sources show a lag of the soft band, it is not obvious that these are due to the same mechanism.

In this letter, we extend these spectral-timing studies to the well-known ULX NGC~1313~X-1, which was recently the subject of an extensive observational campaign with {\em XMM-Newton}, {\em NuSTAR} and {\em Chandra}. The main impetus for this campaign was to study details of the Ultrafast Outflows (UFOs) along our line of sight that were found in archival observations of the source \citep{Pinto2016nature}. Archival observations showed a super-Eddington wind with line-of-sight velocities of 0.2c (which is the escape velocity at 25 Schwarzschild radii). The new observations (Pinto et al., {\em submitted}) show outflows of 0.06c (escape velocity at 150 Schwarzschild radii), suggesting variability in the outflow, possibly due to expansion of the outflow as the accretion rate increases. 

Archival observations of NGC~1313~X-1 have so far shown little variability within individual observations, thus making time lag analysis impossible (e.g. \citealt{Bachetti2013}). However, one orbit in this new campaign caught the source at the end of a flaring episode (see long-term {\em Swift}/XRT light curve in Walton et al., {\em submitted}). In this flaring episode, we detected kilosecond timescale variability, thus allowing time-lag analysis for the first time. In the following letter, we describe the observations in Section~\ref{sec:obs}, the spectral-timing analysis and results in Section~\ref{sec:results}, and discuss in Section~\ref{sec:discuss} how these results compare to time lags in NGC~55~X-1 and NGC~5408~X-1, and how they might fit into a super-Eddington accretion flow framework.

\begin{figure*}
\begin{center}
\includegraphics[width=0.45\textwidth]{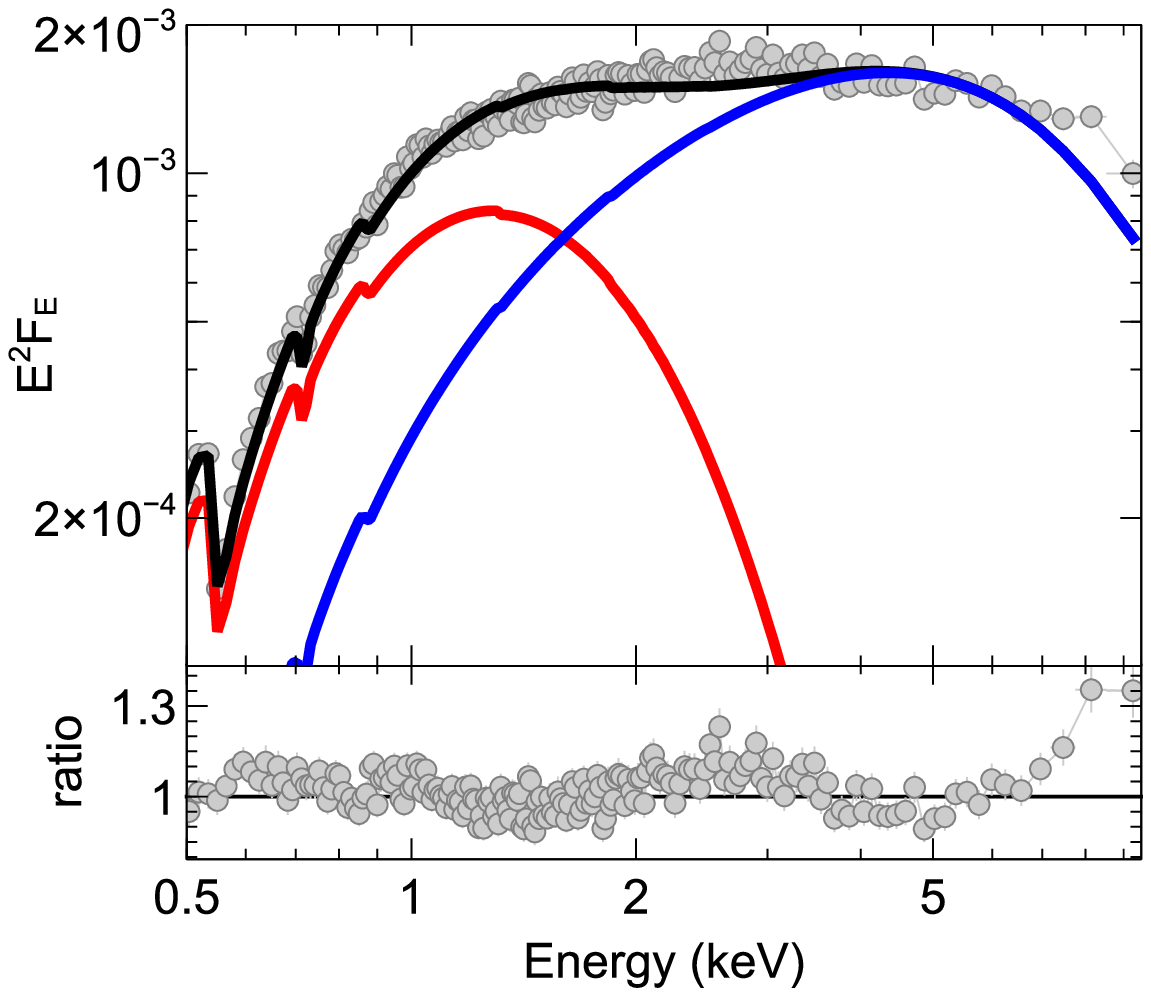}
\includegraphics[width=0.45\textwidth]{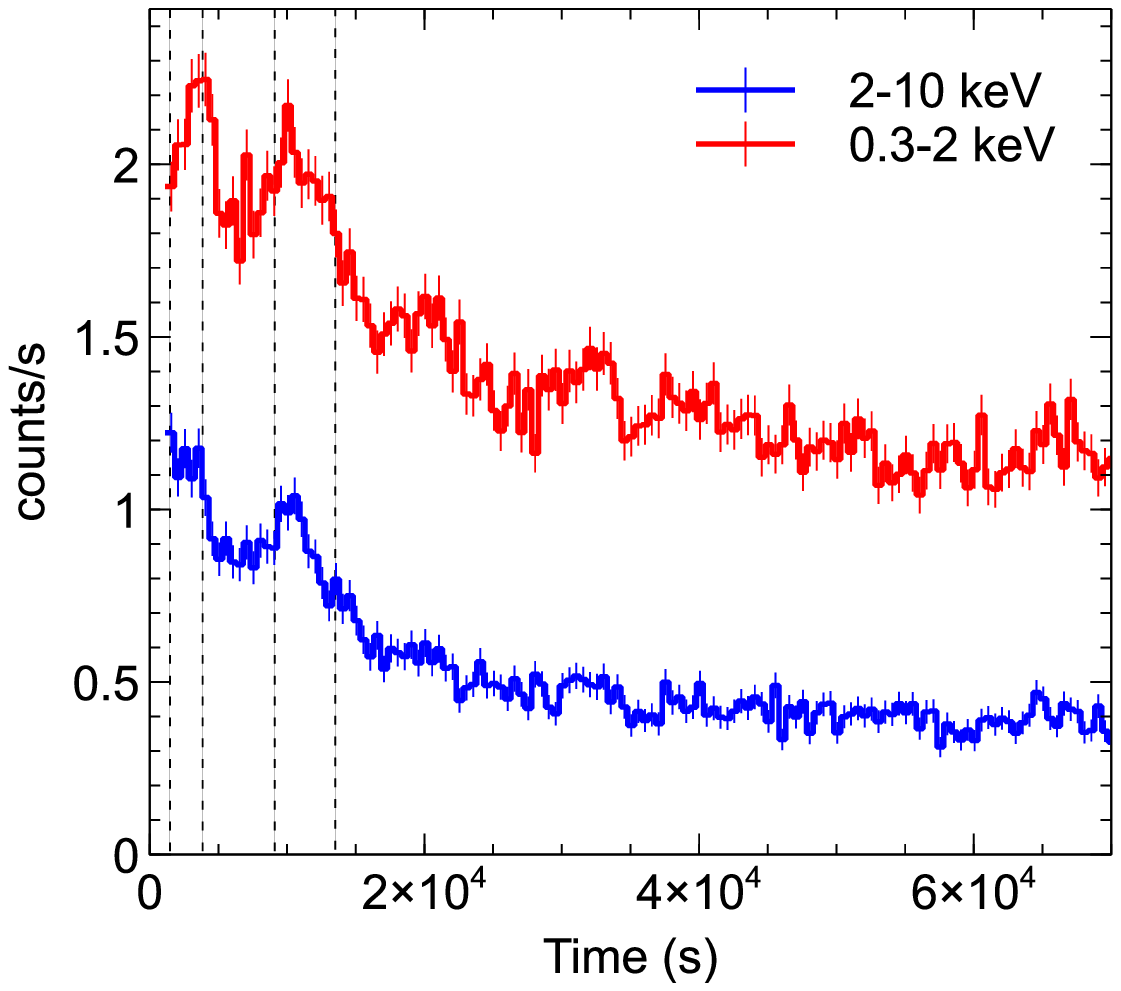}
\caption{{\em Left:} The energy spectrum {\em XMM-Newton} spectrum of OBSID 0803990101, fit with a demonstrative model of two disk blackbody components. 
{\em Right:}  The soft (0.3--2~keV) and hard (2--10~keV) lightcurves from the same observation. 
The vertical dashed lines highlight peaks in the light curve, to guide the eye in comparing between different energy bands.}
\label{fig:lc}
\end{center}
\end{figure*}

\section{Observations and Data Reduction}
\label{sec:obs}
NGC~1313 X-1 was recently observed in an extensive campaign with {\em XMM-Newton} (PI: Pinto), {\em Chandra} (PI: Canizares) and {\em NuSTAR} (PI: Walton) that is described in Pinto et al., {\em submitted} and Walton et al., {\em submitted}. In this letter, we focus on the timing analysis of the {\em XMM-Newton} observations. \textcolor{black}{While all 750~ks of data were analyzed, only the first half of obsid 0803990101  (70~ks) showed sufficient variability }, and thus is the focus of this work. 

We analyze data from the {\em XMM-Newton} EPIC-pn camera \citep{Struder2001} because of its superior effective area and time resolution over the MOS. We reduced the data using the {\em XMM-Newton} Science Analysis System (SAS v. 18.0.0) and newest calibration files. We begin with the observation data files and followed standard procedures. The observations were taken in Full Frame Mode. We produce a circular source extraction region of 35 arcsec in radius, and a background region of 80 arcsec. To avoid periods of high background, we removed time intervals at the beginnings or ends of the observation where the 12--15 keV light curve exceeded 0.4 counts/s. The background was low throughout the remaining observation.
The response matrices were produced using rmfgen and arfgen in SAS. The pn spectra were binned to a minimum of 25 counts per bin to allow for the use of $\chi^2$ statistics for spectral modelling. For the timing analysis, we created background subtracted light curves in different energy bins with the same start and end times with the {\sc SAS} tool {\sc epiclccorr} in 1~second bins.

\begin{figure}	
\begin{center}
\includegraphics[width=\columnwidth]{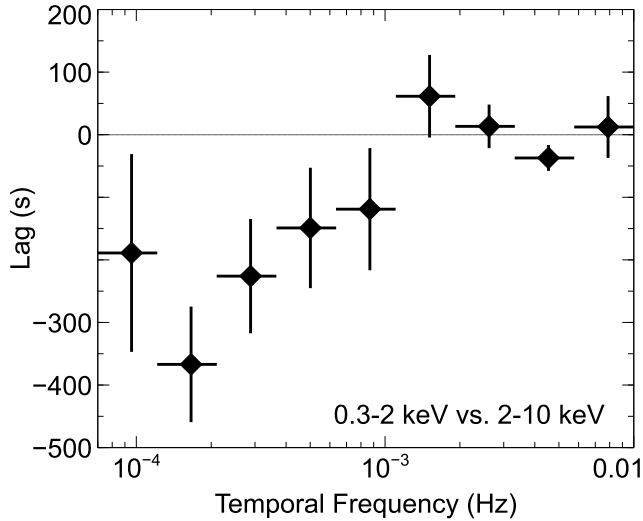}
\caption{The soft (0.3--2~keV) vs. hard (2--10~keV) lags as a function of temporal frequency. By convention, negative lags are soft band lags. The variability  on timescales of thousands of seconds show a soft band lag of $\sim150$~seconds. This is the third ULX to exhibit soft lags.}
\label{fig:lagfreq}
\end{center}
\end{figure}

\section{Results}
\label{sec:results}
Like other ULXs, NGC~1313~X-1 shows a 0.3--10~keV spectrum that is broadly described by two thermal components. We demonstrate this point in Fig.~\ref{fig:lc}-{\em left}, where we show the unfolded 0.3-10~keV {\em XMM-Newton} spectrum fit with a simple phenomenological model ({\sc  tbabs*(diskbb+diskbb)} in XSPEC). See Walton et al., {\em submitted} for a much more thorough analysis of the multi-epoch broadband spectra, including {\em NuSTAR} and {\em Chandra} data. The {\em NuSTAR} data, in particular, imply a third high-energy component that can be interpreted either as due to Compton upscattering in a corona or emission from a neutron star accretion column. This additional hard component also explains the excess seen in the 7--10~keV band in the {\em XMM-Newton} spectrum. 
Focusing now on the {\em XMM-Newton} spectra, one can see from the spectral decomposition shown in Walton et al., {\em submitted} and from Fig.~\ref{fig:lc} here that the soft component contributes mostly below 2~keV, while the hard component dominates above 2~keV. We use this to motivate our search for correlated variability in these two bands. 

\begin{figure}	
\begin{center}
\includegraphics[width=\columnwidth]{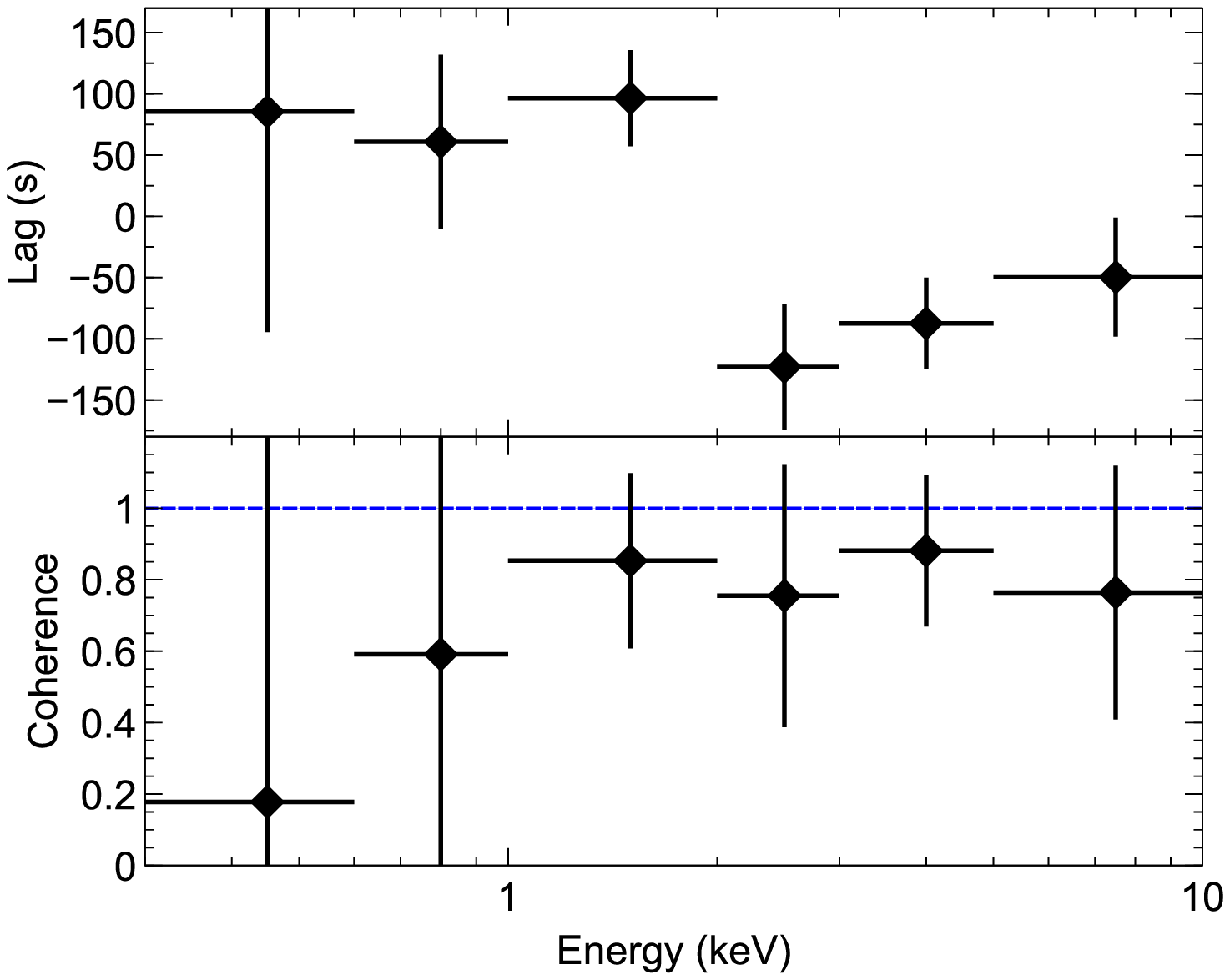}
\caption{The lag-energy spectrum ({\em top}) and coherence spectrum ({\em bottom}) averaged over the frequency range from $[1-5] \times 10^{-4}$~Hz. By convention, the larger the lag, the more delayed the energy bin. The coherence is best constrained above 1~keV.}
\label{fig:lag-coher}
\end{center}
\end{figure}

Fig.~1-{\em right} shows the soft (0.3--2~keV) and hard (2--10~keV) band light curves from the first half of the observation, at the end of a bright flare \textcolor{black}{(See Walton et al., {\em submitted} for all light curves in the campaign)}. Beyond 70~ks, the light curve is constant, and so we focus just on this first half of the observation. \textcolor{black}{The choice to cut the light curve at half the observation length (70~ks) was somewhat arbitrary, and the following results are insensitive to the exact cut-off, as most of the variability occurs in the first $\sim40$~ks.} Variability on kilosecond timescales is clearly present in the data. The vertical dashed lines guide the eye, and one can begin to see a slight delay of the soft band with respect to the hard.

To confirm and quantify this delay, we performed a Fourier timing analysis, by measuring the frequency-dependent time lag between the soft band (defined as 0.3--2~keV) and the hard band (2--10~keV), following the procedure outlined in \citet{uttley14}. Briefly, we take the Fourier transform of the two light curves (0.3--2~keV and 2--10~keV). The frequency-dependent time lag is related to the phase difference between the Fourier transforms of these two complete light curves, binned from $7\times 10^{-5}$~Hz to $10^{-2}$~Hz in equal logarithmic-spaced bins, so that error are Gaussian. We observe a soft band lag of $140\pm40$~seconds over the frequency range $[1-5] \times 10^{-4}$~Hz (Fig.~\ref{fig:lagfreq}). 

\begin{figure}	
\begin{center}
\includegraphics[width=\columnwidth]{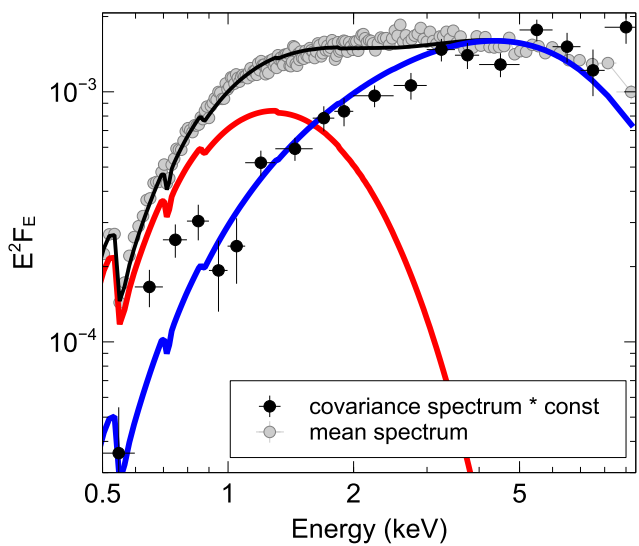}
\caption{The covariance spectrum (black) computed over the frequencies $[1-5] \times 10^{-4}$~Hz, compared to the time-integrated energy spectrum (gray; same as Fig.~\ref{fig:lc}). The covariance spectrum (i.e. the spectrum of the emission that is linearly correlated to the broad 0.3--10~keV reference band) has been scaled by an arbitrary factor of 3 in order to more clearly compare the shapes of these two spectra. The covariance spectrum is much harder than the time-integrated spectrum, and follows the shape of the hotter disc blackbody component (blue), and not the lower temperature component (red).}
\label{fig:covspec}
\end{center}
\end{figure}

We next compute the lag-energy spectrum and the coherence averaged over the frequencies where we observe the soft lag, roughly $[1-5] \times 10^{-4}$~Hz (Fig.~\ref{fig:lag-coher}). The lag/coherence is computed between each bin and a broad reference band (0.3--10~keV with the bin-of-interest removed). The coherence measures the degree to which each bin is a simple linear transformation of the broad reference band, normalized to 1 \citep{nowak99}. The coherence is high above 1~keV, and the lag shows a fairly sharp drop above 2~keV. 

Probing further, we computed the frequency-dependent covariance spectrum (see, e.g. \citealt{wilkinson09}), which is the component of the energy spectrum that is varying coherently. The covariance is similar to the rms spectrum, but specifically shows the emission that is linearly correlated with the 0.3--10~keV reference band, and therefore highlights the spectral component that contributes to the observed time lags. In Fig.~\ref{fig:covspec}, we compare the covariance spectrum (over the frequency band where the soft lag is observed: $[1-5] \times 10^{-4}$~Hz) compared to the time-averaged energy spectrum (also shown in Fig.~\ref{fig:lc}-{\em left}). Because the covariance spectrum is measured over a specific frequency range, its normalization will always be less than the time-averaged energy spectrum, and so we have scaled the covariance spectrum by an arbitrary factor of 3 in order to better compare the shapes of the two spectra. It is clear that the covariance spectrum is harder than the mean spectrum. The covariance spectrum seems to follow the shape of the hotter blackbody component, and there is little evidence for a low-temperature component (similar to results found in \citealt{middleton15} for a sample of ULXs). While there are qualitative similarities between the hard component of the mean spectrum and the covariance spectrum, the blackbody temperature for the covariance spectrum is {\em slightly} hotter (kT = $2.28\pm0.25$) than that required by the mean spectrum (kT = $1.91\pm0.07$; blue line in Fig.~\ref{fig:covspec}). The covariance spectrum provides independent confirmation that there are two distinct components below 10~keV and suggests that it is the hotter component that is responsible for the lag shown in Fig.~\ref{fig:lagfreq}. 

A few features become apparent when comparing the mean spectrum to the covariance. First, there is an apparent excess in the 7--10~keV band in the mean spectrum and especially in the covariance spectrum. This is likely because we have not accounted for the hard non-thermal component that is seen clearly in the {\em NuSTAR} spectra (Walton et al., {\em submitted}). That the excess is also seen in the covariance spectrum may suggest that the non-thermal component varies coherently with the hotter disk blackbody component, or simply that the hotter component is broader than a simple disk blackbody component  (e.g. a Comptonization model, as in {\sc CompTT} or a blackbody where radial advection is important, as in the {\sc Diskpbb} model; Walton et al., {\em submitted}). 
Second, there is a tentative excess in the covariance at 0.6-0.9~keV. We note that this excess is around the energy of the OVIII and FeXVII emission lines found in \citet{Pinto2016nature}, and also Pinto et al., {\em submitted}.

\section{Discussion}
\label{sec:discuss}

During the recent {\em XMM-Newton} campaign of NGC~1313~X-1, one observation caught the end of a flaring period. This allowed us to perform a detailed spectral-timing and time lag analysis, revealing the following:
\begin{itemize}
    \item The broadband spectrum is broadly described by two blackbody-like components. The soft component dominates from 0.3--2~keV, and hard from 2--10~keV.
    \item The light curve varies on timescales of kiloseconds, and on these timescales, the soft band is observed to lag the hard by $\sim 150$~s.
    \item The covariance spectrum (i.e. the emission that contributes to the lags) is harder than the time-averaged spectrum, and resembles the shape of the hotter thermal component.
\end{itemize}

A 150-second lag in a stellar-mass compact object is a very long timescale---closer to the timescales measured in AGN than typical X-ray binaries. There is no clear consensus on its origin. The resemblance of the covariance spectrum to the hotter component suggest that processes near the inner accretion flow are responsible for the lag. Interestingly, this was also noted by \citep{hernandez-garcia15} for NGC~5408~X-1 for timescales that are roughly 100 times shorter than those seen here in NGC~1313~X-1. In the subsections below, we discuss possible origins of the lag and compare to other ULXs.

\subsection{The origin of the lag}

The lack of high-frequency variability in most ULXs has been noted by several authors (e.g. \citealt{heil09,Bachetti2013}), leading to the suggestion that short timescale variability is suppressed due to reprocessing in the enveloping accretion flow \citep{middleton15,mushtukov19}. This could also explain soft band lags, as hotter inner disc photons Compton scatter off cooler electrons in the outflow \citep{Middleton2019}. 

If our $\sim 150$~second soft lag is the timescale of Compton scattering through a medium confined to within the spherization radius $R_{\mathrm{sp}}$, it allows us to \textcolor{black}{put a lower limit on} the density of the outflow. We start by assuming a medium whose dominant source of opacity is Compton scattering (as in \citealt{mushtukov19,Middleton2019}). We assume the effective optical depth is $\tau_e = \sigma_T n_e R_{\mathrm{sp}}$, where $\sigma_T$ is the Thomson scattering cross section and $n_e$ is the number density of free electrons. The timescale for photons to escape the medium ($t_{\mathrm{sc}}$) is set by the electron mean free path $l_{\mathrm{sc}}=1/\sigma_T n_e$ and number of scatterings $n$, such that $t_{\mathrm{sc}}=n (l_{\mathrm{sc}}/c)$. Given that the number of scatterings scales roughly as the square of the optical depth \citep{Rybicki79}, we can rewrite the timescale for Compton scattering as
$$ t_{sc} = \sigma_T n_e R_{\mathrm{sp}}^{2}/c. $$

A $10 M_{\odot}$ compact object accreting at $\dot M = 10 \dot M_{\mathrm{Edd}}$ will have a spherization radius of roughly $R_{\mathrm{sp}} \sim \dot M R_{\mathrm{in}} \sim 100~r_{\mathrm{g}}$, where we conservatively assume an inner disc radius of $R_{\mathrm{in}}=10 r_{\mathrm{g}}$. It follows that for a 150~second Compton scattering timescale, the electron number density is $n_e = 3 \times 10^{20}$~cm$^{-3}$. This inferred density is extremely high for a wind, and is more consistent with the midplane density seen in magnetohydrodynamical simulations of super-Eddington accretion flows around stellar mass black holes \citep{sadowski16}. It implies a large number of scatterings ($\sim 10^{4}$), which would effectively thermalize the wind, preventing further Compton downscattering.  It is likely that a medium cool enough to downscatter few keV photons would have non-negligible bound-free opacity over these path lengths.

This high density issue is exacerbated if the time lag is produced in the inner accretion flow ($<100 r_{\mathrm{g}}$), as suggested by the resemblance of the covariance spectrum to the hotter disc component.  Moreover, NGC~1313~X-1 also shows hard X-ray component up to 20~keV (Walton et al., {\em submitted}). If this is associated with Compton up-scattering (rather than from a neutron star accretion column), it requires very different electron temperatures in a confined region. Perhaps this last point can be reconciled if this hardest component is produced via bulk motion Comptonization in the outflow itself \citep{soria11}, where there is a faster spine that  can up-scatter inner disc photons to 20~keV, and a slower sheath that down-scatters photons to produce the observed lag. 

\citet{middleton15} noted that in addition to extrinsic variability due to scattering in the wind, there may also be signatures of intrinsic variability in the inner accretion flow (often called `propagating fluctuations' \citealt{Lyubarskii97,kotov01}), where mass accretion rate fluctuations propagate inwards on the viscous time. In X-ray binaries and AGN this produces a hard lag (e.g. \citealt{nowak99}, \citealt{papadakis01}), not a soft lag, as is seen here for NGC~1313~X-1. While our observed lags do not resemble the hard continuum lags nearly ubiquitously seen in sub-Eddington black holes, we cannot rule out that these lags are intrinsic to the continuum as the disc structure in this super-Eddington system is very different. For instance (and rather speculatively), reverse shocks propagating out through a large scale height inner disc could produce soft lags, rather than hard. Soft band lags that are intrinsic to super-Eddington accretion flows should be explored further, but are beyond the scope of this letter.

\begin{figure}	
\begin{center}
\includegraphics[width=\columnwidth]{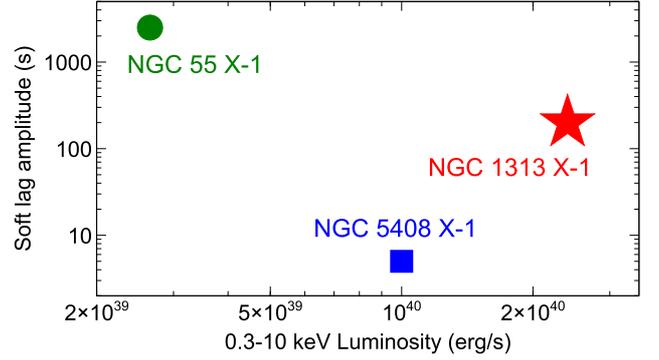}
\caption{Comparison of the time lags seen here in NGC~1313 X-1 to the two other ULXs with measured lags, NGC~55 X-1 (lag amplitude and unabsorbed luminosity taken from \citealt{Pinto2017}) and NGC~5408 X-1 (lag amplitude from \citealt{demarco13} and unabsorbed luminosity from \citealt{miller13}). Note the log scale on the x-axis.}
\label{fig:discuss}
\end{center}
\end{figure}

\subsection{Comparison to other ULXs}

Time lags have been measured in three ULXs thus far: NGC~5408~X-1 \citep{heil10,demarco13,hernandez-garcia15}, NGC~55~X-1 \citep{Pinto2017}, and now in NGC~1313~X-1. All three of these sources exhibit soft lags, but the variability timescales and the amplitudes of the lags are very different. In Fig.~\ref{fig:discuss}, we plot the lag amplitude vs. luminosity from the literature compared to our results here. There does not appear to  be any correlation with luminosity or with spectral hardness, and given the vast differences in lag timescales, it is not clear at this point that the lags are produced by the same mechanism. 

One tantalizing observation is that, while the lag amplitudes differ by up to three orders of magnitude, they are each about a few to a few tens of percent of the characteristic variability timescale for each source, which suggests a degree of commonality of origin. NGC 1313 X-1 and NGC 5408 X-1 both show smaller lags per characteristic timescales than NGC 55 X-1 and also show similarities in their covariance spectra. For NGC 1313 X-1, the soft lag amplitude is $\sim 5$ per cent of the average variability timescale (150s lag amplitude over timescales of 2--10 ksec). Similarly, the soft lag in NGC 5408 X-1 is $\sim 5$ per cent of its average variability timescale. These two ULXs also show covariance spectra that are harder than the mean, time-integrated energy spectrum \citep{hernandez-garcia15,middleton15}. NGC 55 X-1, on the other hand, shows a soft lag that is 10-30 per cent of the variability timescale, which is close to, but clearly larger than, the lag fractions of the other two. Its covariance spectrum shows little difference from the mean spectrum (other than normalization; \citealt{Pinto2017}). Perhaps this suggests that all three share common origins but that both the low- and high-temperature components contribute to the lags in NGC 55 X-1, whereas in the other two, more luminous ULXs, the lags are produced predominantly by the hotter component.

NGC~5408~X-1 shows the  highest frequency soft lag that occurs at the same frequency as a broad quasi-periodic oscillation (QPO; \citealt{heil10,demarco13,li17}). No such QPOs are found at the timescales of the soft lags in NGC~55~X-1 or here in NGC~1313~X-1, although in NGC~1313~X-1 the lag occurs over a broad frequency range, and so we cannot rule out that there is a broad Lorentzian in the power spectrum over this frequency range. NGC~5408~X-1 also shows evidence for a low-frequency hard lag \citep{hernandez-garcia15}. While we find no evidence for low-frequency hard lags in the other two sources, this may be due to the inability to probe long enough timescales. Given these differences, we caution the reader about associating all soft lags in ULXs with the same mechanism, but given their relative novelty, we allow ourselves, in the following section, to speculate on such a unified picture. 

\subsection{Time lags in the context of a unified super-Eddington outflow model}

Several authors have proposed a unified picture for super-Eddington accretion in ULXs, \textcolor{black}{whereby all systems have roughly the same mass (e.g. $\sim 10~M_{\odot}$), just viewed at different inclinations}. \citet{Poutanen2007} first suggested that softer, less luminous ULXs are viewed at a larger inclination, and therefore our line-of-sight to the innermost regions is blocked by the large scale height wind \textcolor{black}{(see also \citealt{dauser17} for Monte-Carlo simulations based on this geometry)}. This unified picture was supported observationally by \citet{Sutton2013}, who demonstrated the inclination-dependence of the spectra and the soft and hard band excess variance. The rms variability \citep{heil09} and covariance spectra \citep{middleton15} also fit into this global picture.  Recently, \citet{pinto19} proposed that the observed UFOs in ULXs also support this unified picture, where harder sources  (presumably viewed more face-on), showed higher velocity outflows, as our line of intercepts the wind at smaller radii, closer to the launching radius. 

Previous attempts to unify ULX in a UFO context suggest that NGC~55~X-1, with its lower luminosity, softer spectrum and slower winds, may be oriented  at a larger viewing angle than NGC~1313~X-1 and NGC~5408~X-1. If the emission from high-inclination objects originates from a larger region, it is feasible that the time lags (regardless of the interpretation) would be larger than those viewed at lower inclinations. 
In this simple inclination-dependent scenario, we would expect NGC~1313~X-1 (the hardest and most luminous of this sample) to have the smallest lag, which is not the case. Perhaps this is due to a further dependence on mass accretion rate (e.g. \citealt{Feng2016}). For NGC~1313~X-1, we only observed the soft lags during a flaring state. The spectral modelling in Pinto et al., {\em submitted} suggest that when the flux is high and spectrum soft, the X-ray emitting region and the wind launching radius are larger than usual. In a unified picture for the time lags that depends on both inclination and mass accretion rate, this could explain why we observe larger lags in NGC~1313~X-1 compared to NGC~5408~X-1.

\section{Conclusions}

We present the discovery of a $\sim150$~second soft lag in the ULX NGC~1313~X-1. 
While the origin of soft lags in ULXs is not well-understood, the covariance spectrum suggests that this is due mostly to processes in the hotter component, thought to originate in the inner accretion flow. Thus far, all ULXs that show soft lags, also show evidence for relativistic outflows through measurements of narrow atomic features in both emission and absorption. It is not yet clear if this is an observational bias (as both time lag analysis and {\em XMM-Newton}/RGS spectral analysis require high signal-to-noise observations) or if this is suggesting that processes in the outflow are responsible for the lag. If it is the latter, then time lags can provide a new tool for constraining the density and extent of radiatively-driven outflows. Monitoring of ULXs to catch them in variable states, together with deep follow-up observations will put this idea to the test.

\section*{Acknowledgements}
Thank you to the referee Thomas Dauser for helpful comments that improved this work. EK acknowledges support from NASA Award Number 80NSSC19K0176. CP is supported by ESA Research Fellowships. TPR thanks STFC for support as part of the consolidated grant ST/K000861/1. CRC acknowledges support by the Smithsonian Astrophysical Observatory (SAO) contract SV3-73016 to MIT, which is in turn supported by NASA under contract NAS8-03060.

\bibliographystyle{mnras}
\bibliography{bibliografia}

\end{document}